\documentclass{caosp309}
\usepackage{graphicx}
\usepackage{natbib}
\articleNo{XXX}
\pubyear{2025}
\volume{55}
\volnumber{1}
\firstpage{1}
\received{Nov 27, 2024}
\accepted{Dec 31, 2024}

\begin{document}
\hauthor{H.~M.~J. Boffin \& D. Jones}
\title{The importance of binary stars}
\author{Henri M.J. Boffin\inst{1}\orcid{0000-0002-9486-4840}
\and
David Jones\inst{2,3,4}\orcid{0000-0003-3947-5946}}
\institute{ESO, Karl-Schwarzschild-str. 2, 85748 Garching, Germany \email{hboffin@eso.org}
\and
Instituto de Astrof\`\i sica de Canarias, 38205 La Laguna, Tenerife, Spain \email{djones@iac.es}
\and
Departamento de Astrof\`\i sica, Universidad de La Laguna, E-38206 La Laguna, Tenerife, Spain
\and
Nordic Optical Telescope, Rambla Jos\'e Ana Fern\'andez P\'erez 7, 38711, Bre\~na Baja, Spain}
\date{November 27, 2024}

\maketitle

\begin{abstract}
Stars are mostly found in binary and multiple systems, as at least 50\% of all solar-like stars have companions — a fraction that goes up to 100\% for the most massive stars. Moreover, a large fraction of them will interact in some way or another over the course of their lives. Such interactions can, and often will, alter the structure and evolution of both components in the system. This will, in turn, lead to the production of exotic objects whose existence cannot be explained by standard single star evolution models, including gravitational wave progenitors, blue stragglers, symbiotic and barium stars, novae, and supernovae.
More generally, binary stars prove crucial in many aspects, ranging from cultural ones, to constraining models of stellar evolution, star formation,  and even, possibly, of gravity itself. They also provide a quasi-model independent way to determine stellar masses, radii, and luminosities. We here provide a brief summary of the importance of binary stars. 

% I will lead a walk in the zoo of binary stars, highlighting some specific examples and why binary stars are important.

\end{abstract}

\section{Binary stars are ubiquitous}
The Sun is a single star -- which is likely one of the reasons we are here (see Sec.~\ref{Sec:Pla}) -- but a large number of stars, and the majority of solar-like and massive stars, are members of binary systems \citep{2015MNRAS.448.1761W,2017ApJ...836..139F,2024BSRSL..93..170M}, with the mass ratio distribution depending on the primary mass \citep{1993A&A...271..125B,2019MmSAI..90..359B,2020Obs...140....1B}. Looking at the night sky, its brightest star, Sirius, is a binary. Discovered by \citet{1844MNRAS...6R.136B}, it is reported in a paper of which we cannot resist quoting the first sentence as it is still most relevant when discussing binary stars:
\begin{quote}
The subject which I wish to communicate to you, seems to me so important for the whole of practical astronomy, that I think it worthy of having your attention directed to it.
\end{quote}
The companion of Sirius, discovered via astrometry, turned out to be a white dwarf (WD), and Sirius is now the prototype of its own class of systems 
\citep{2013ApJ...769....7L,2017ApJ...840...70B}. Such systems, containing a main-sequence star and a WD, can also be used to infer the ages of the main-sequence stars \citep{2019ApJ...870....9F,2023MNRAS.524..695Z}.

\begin{figure}
\centerline{\includegraphics[width=\textwidth]{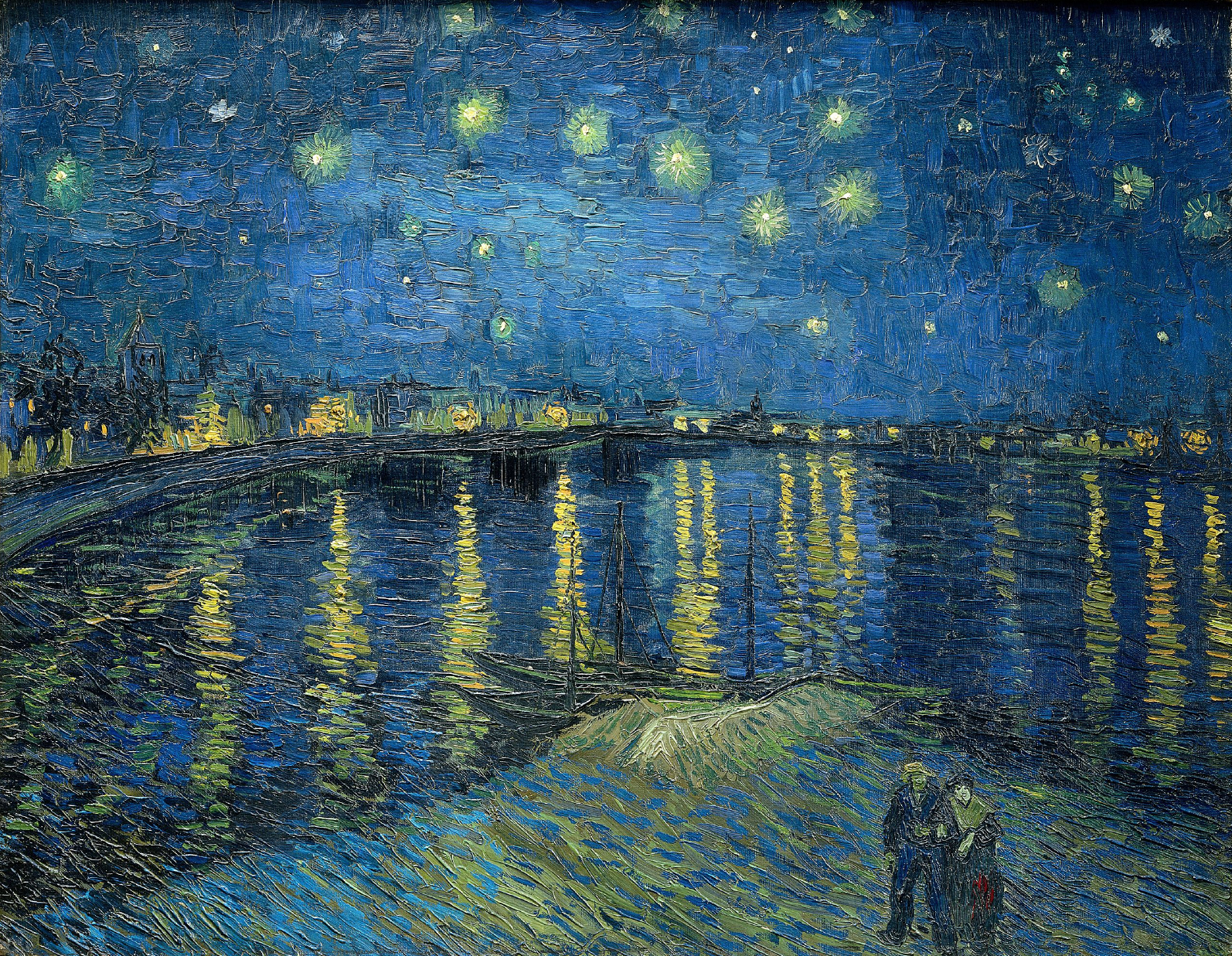}}
\caption{Starry Night Over the Rhone by Vincent Van Gogh, showing the Big Dipper and the binary system Mizar and Alcor. Oil on canvas, Musée d'Orsay, Paris.}
\label{fig:VanGogh}
\end{figure}

In our close vicinity, we also find binary stars. Thus, Alpha Centauri is a binary system of solar-like stars, located just 4.3 light-years from the Sun, and appearing as a visual binary with an orbital period of about 80 years \citep{1982Obs...102...42H,2016A&A...586A..90P}. 
In fact, it is likely that Alpha Centauri forms with the closest star to the Sun, Proxima Centauri, a triple system! 
Another of the closest systems to be Sun, located only 2 parsecs away, is the binary brown dwarf that forms Luhman 16 \citep{2013ApJ...767L...1L,2014A&A...561L...4B,2017MNRAS.470.1140B,2024AN....34530158B}. 
Finally, another famous multiple is the couple Mizar and Alcor in the Big Dipper (Fig.~\ref{fig:VanGogh}), known as {\it jumyouboshi} (or ``lifespan star'') in Japan because it is thought that if you cannot distinguish both stars with your naked eyes, you will die within a year. Astronomy can be deadly! We now know that Alcor and Mizar are themselves binaries, while each components of Mizar is another binary \citep{2010AJ....139..919M}. Thus, we have here nice examples of binaries on various scales and discovered by various techniques. A rather similar bright system of six stars in three binary systems is given by Castor, the second brightest star in the constellation Gemini. 

\section{Binary stars constraint stellar models}
As shown by the Alcor \& Mizar example, binary stars span a wide range of separations. At one extreme, we have the binary star HM Cancri, which has the shortest orbit of all the known AM CVn systems. These are composed of a WD accreting from another WD or Helium star. As such, they have all short orbital periods, with the one of HM Cancri being 321 seconds \citep{2002A&A...386L..13I,2010ApJ...711L.138R}. This system is a nice laboratory as  \cite{2023MNRAS.518.5123M} show that mass transfer is counteracting gravitational-wave dominated orbital decay and that it is one of the brightest verification binaries for the Laser Interferometer Space Antenna spacecraft. On the other extreme, we have common proper motion pairs, which are weakly gravitationally bound with orbital periods of up to several millions of years.

One of the principal uses of binary stars is to obtain accurate radius, masses, and luminosities of stars. This will then be very useful to constrain stellar evolution models. The best systems for this are double-eclipsing binaries with spectroscopic and photometric orbits, as in this case, one can obtain all the parameters, in an almost model-independent way. Some of the most recent examples can be found in the series of papers by J.-L. Halbwachs and colleagues \citep[for example,][]{2020MNRAS.496.1355H} and those by John Southworth in {\tt The Observatory}. In \cite{2024arXiv240419443S}, he determines the masses and radii of the component solar-like stars of the binary system V454 Aurigae with amazing precision\footnote{The values in his Table III are in fact even more precise that what is written in the abstract and is quoted here.}, finding masses of 1.034 $\pm$ 0.006 M$_\odot$ and 1.161 $\pm$ 0.008 M$_\odot$, and radii of 0.979 $\pm$ 0.003 R$_\odot$  and 1.211 $\pm$ 0.003 R$_\odot$. If these values are also as accurate, then they are wonderful tests of theoretical models, although it would be useful to determine the abundance of elements in the system with great accuracy as well for even more constraints. 
More such systems are listed in the Detached Eclipsing Binary Catalogue\footnote{\url{https://www.astro.keele.ac.uk/jkt/debcat/}}, while another interesting example that could serve as a test for magnetic stellar evolution models can be found in \cite{2024A&A...687A.116H}. 
This can also be done with wide binaries, especially those that contain a WD and a MS star \citep{2019ApJ...870....9F,2021MNRAS.505.3165R,2022ApJ...929...26M}.

\begin{figure}
\centerline{\includegraphics[width=\textwidth]{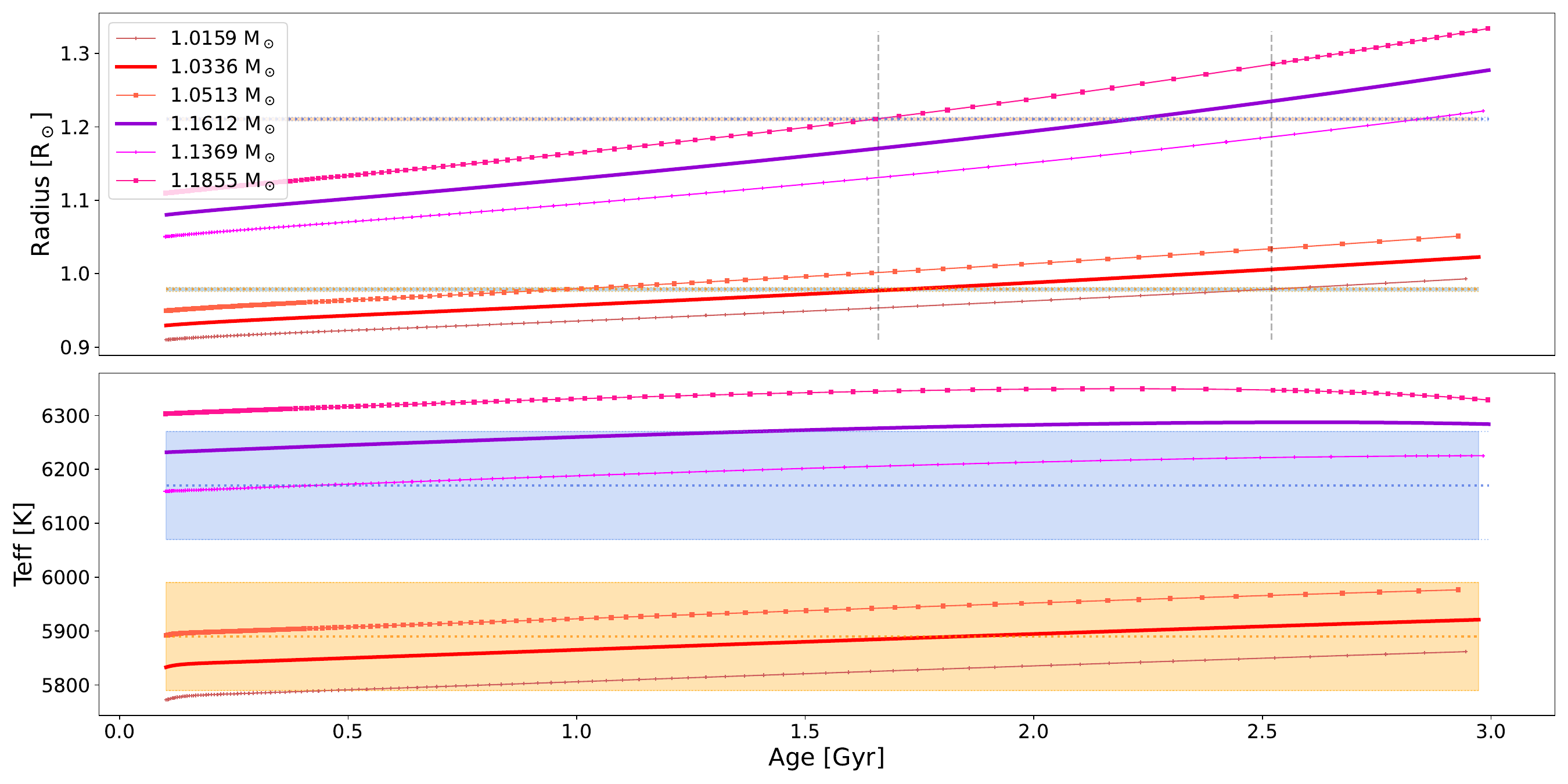}}
\caption{Comparison between MIST models for solar metallicity stars of masses corresponding to the components of V454 Aurigae and the  parameters determined by  \cite{2024arXiv240419443S}. The top panel shows the radius as a function of age, while the bottom panel concerns the effective temperature as a function of age. In each panel, the lower set of curves correspond to the less massive component, the upper ones to the more massive one. The determined values of the radii are so precise that the allowed range appears like horizontal lines, while for the temperatures, the uncertainty is greater and the possible ranges are shown with the shaded areas.}
\label{fig:masses}
\end{figure}

\begin{figure}
\centerline{\includegraphics[width=\textwidth]{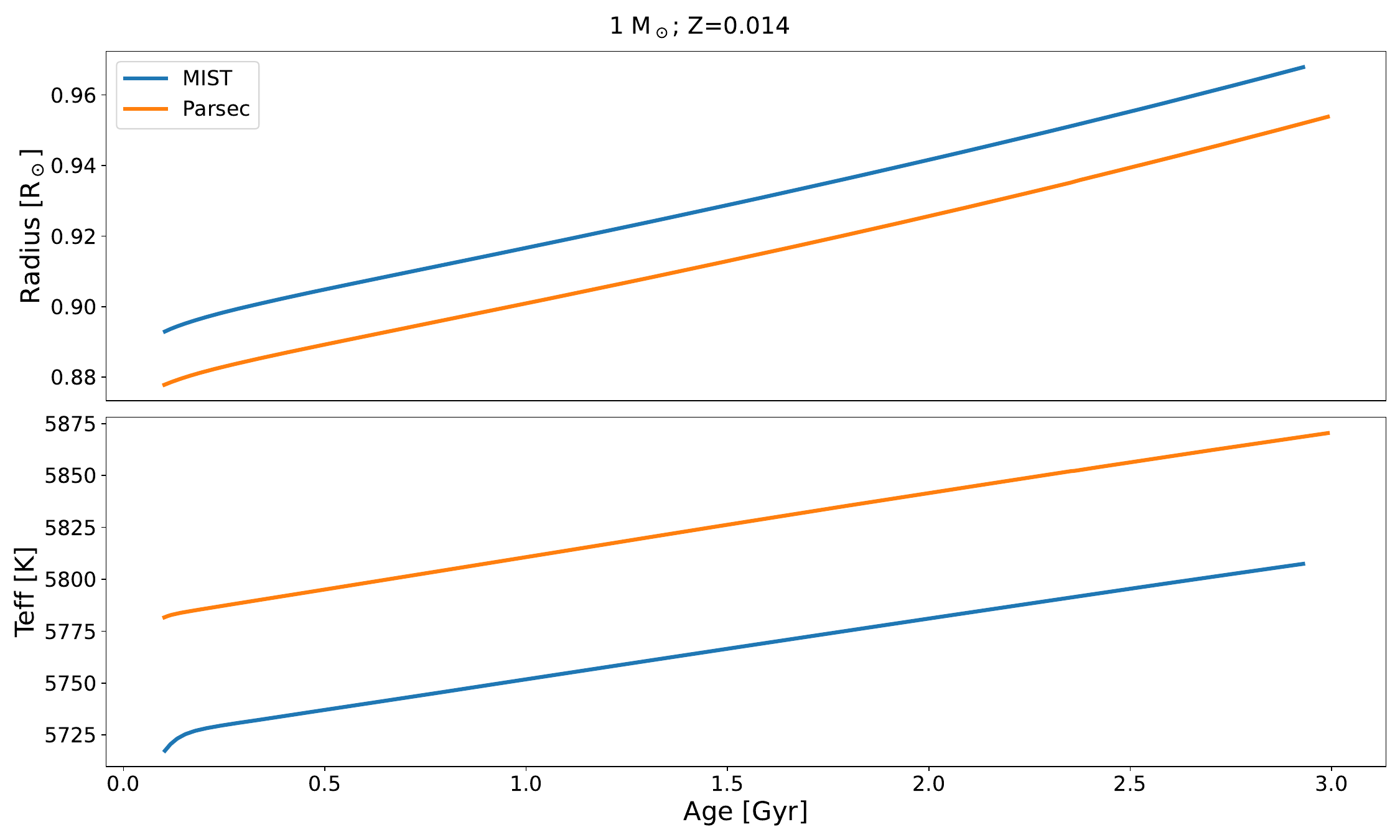}}
\caption{Comparison between MIST and PARSEC evolutionary tracks for a solar mass star of Z=0.014. The panels show the same as in Fig.~\ref{fig:masses}: top is the radius as a function of age; bottom is the effective temperature.}
\label{fig:tracks}
\end{figure}

The data on V454 Aurigae can in principle be used to infer, in a model-independent way, the ages of the stars, assuming both components are coeval -- a reasonable assumption -- and thus test stellar models. 
Figure~\ref{fig:masses} shows a comparison between the precise radii and less precise temperatures of both components in V454 Aurigae with MIST\footnote{\url{https://waps.cfa.harvard.edu/MIST/}} stellar evolution models, assuming solar metallicity and masses in the range determined by \cite{2024arXiv240419443S}. As can be seen, the precise radius allows us in principle to determine the most plausible interval of ages (1.66 to 2.33 Gyr), while the agreement is not so good for the temperature of the stars, especially as far as the most massive component is concerned. The quality of the data we can now achieve will likely also allow us to constrain the assumptions used by different stellar evolution models. In  Fig.~\ref{fig:tracks}, we compare the evolution of the radius and the temperature as a function of age for a Sun, using both MIST and PARSEC\footnote{\url{http://stev.oapd.inaf.it/PARSEC/}} evolutionary tracks. The difference is remarkable and probably due to different calibrations.  When the components of the binary system are pulsating, one can then combine asteroseismology with mass estimates, thereby constraining ages. This is detailed in the contribution of Simon Murphy in this volume. 

In the previous paragraph, we mentioned that the high-quality data provide constraints in a model-independent way. This is not fully correct. Indeed, the analysis of light curves at the precision done in such examples will depend to some extent on the limb-darkening law  -- in this case, a Power-2 one -- and the coefficients used in the law. These are in fact derived from models \citep{2023A&A...674A..63C}! Thus, it is important to ensure self-consistency when doing such an analysis. 

One can also constrain the mass-luminosity relation using binaries and compare with models, as done for example in \cite{2019ApJ...871...63M} or in  \cite{2024MNRAS.528.4272H}, where the latter paper also highlights differences between MIST and PARSEC tracks. Similarly, binary stars containing a white dwarf can be used to constrain models as far as the initial-final mass of white dwarfs are concerned \citep{2024MNRAS.527.9061H}, although one needs to be careful that the white dwarfs are not the result of mergers in case the binary was initially a triple system \citep{2022ApJ...934..148H,2024ApJ...969...68H}. 
A thorough discussion of the impact of binary stars on stellar evolution can be found in the book edited by \cite{Beccari_Boffin_2019}.

\begin{figure}
\centerline{\includegraphics[width=\textwidth]{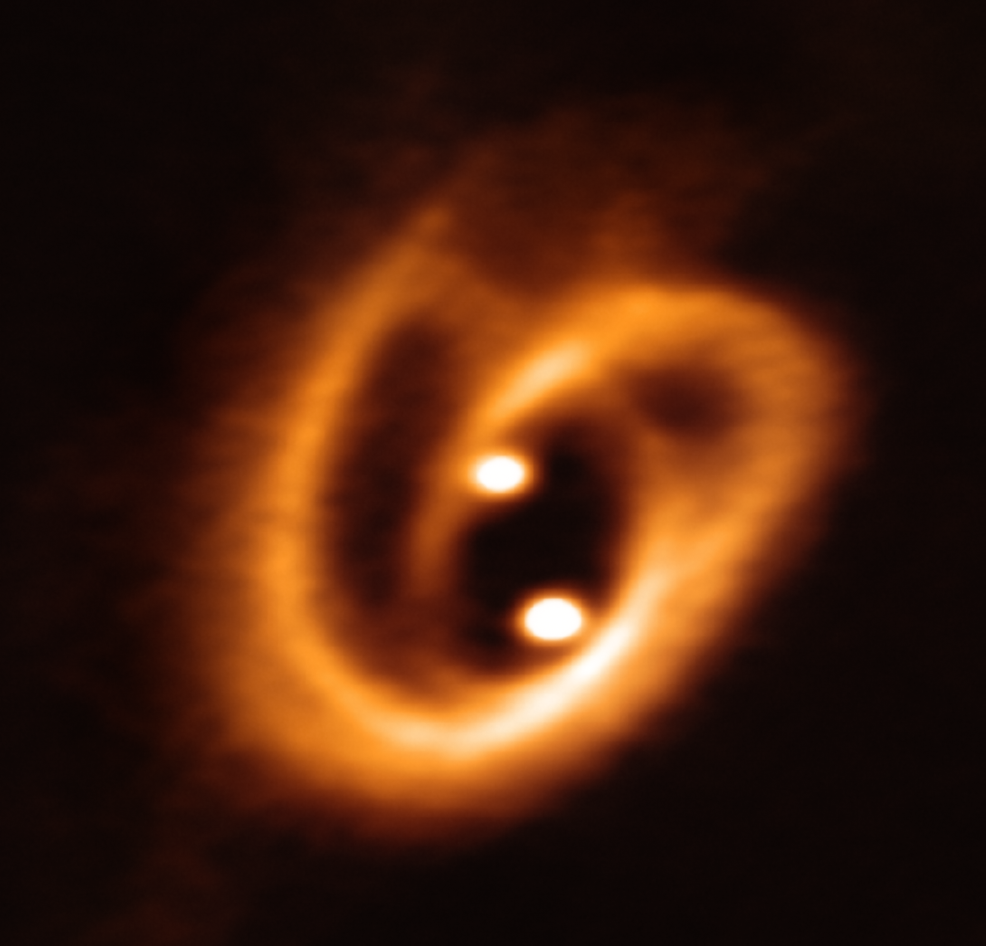}}
\caption{Some stars form as binaries, as shown in this rendering based on ALMA data. 
Credit: ALMA (ESO/NAOJ/NRAO), \cite{2019Sci...366...90A}}
\label{fig:starform}
\end{figure}

\section{Binary stars constraint theories}
Many stars form as binaries (Fig.~\ref{fig:starform}) or multiple stars. Thus, binaries can also tell us a lot about the formation of stars. Recent work has shown that the fraction of binaries is anti-correlated with metallicity at short separation: metal-poor stars appear to present a multiplicity fraction a factor of 2-3 higher than metal-rich stars \citep{2018ApJ...854..147B}. %—> indication that most binaries form by disc fragmentation - Badenes+ 18; Moe+ 19; El-Badry \& Rix 19
In addition, \cite{2020MNRAS.499.1607M} found another anti-correlation, even steeper than the one with metallicity, this time between the $\alpha$-elements and binarity. Other parameters such as mass or age seem to have a smaller impact on the close binary fraction. They interpret this as due to the $\alpha$-elements being trapped in dust and ices, and this is thus another indication that stars form via disc fragmentation.

Binary stars can also be used to test gravity theories. The discovery of binary pulsars and how their orbital period decays was a remarkable confirmation of General Relativity \citep{1975ApJ...195L..51H,1982ApJ...253..908T,2024LRR....27....5F}. More recently, this was also illustrated with the discovery of gravitational waves emitted by the in-spiralling of binary neutron stars or binary black holes \citep{2016PhRvL.116f1102A,2017PhRvL.119p1101A,2023PhRvX..13a1048A}. The recent discovery of BH3 in unreleased Gaia data -- the third example of a quiescent black hole (BH) in a binary system -- is quite interesting as, with an estimated mass of 33~M$_\odot$, it appears to be the most massive BH in the Milky Way and is therefore similar to those discovered from their gravitational wave emissions \citep{2024A&A...686L...2G}. It appears to have a metallicity of $-2.56$, indicating that massive BHs are the remnants of metal-poor stars and are likely formed through dynamical exchanges in clusters \citep{2023MNRAS.526..740R,2024MNRAS.527.4031T,2024A&A...688L...2M}, although alternative theories exist \citep{2024A&A...690A.144I}.

Extremely wide binary stars, with weak gravitational interaction between both components, are in principle a good test of gravitation theories. Indeed, if their separation is larger than roughly 10,000 au, then the MOND theory of gravitation predicts higher orbital velocities than those expected with Newtonian dynamics \citep{2012EPJC...72.1884H}. Many studies have been done in recent years, following the Gaia data releases, including \cite{2019MNRAS.488.4740P,2023MNRAS.525.1401H,2023ApJ...952..128C,2024ApJ...960..114C,2024MNRAS.527.4573B,2024ApJ...972..186C,2024MNRAS.533..110C} and \citet{2024arXiv241017178H}.
However, such measurements prove difficult as the external field effect predicted by MOND is less severe in the Galactic plane, where most such binary stars can be studied, than in the Halo. Moreover, one needs to have all three components of the relative velocities and should correct for projection effects \citep{2019MNRAS.482.5018E,2024NewAR..9801694E}. For now, therefore, it seems like the results are still inconclusive. It is clear, however, that they could have a tremendous impact.

\begin{figure}
\centerline{\includegraphics[width=\textwidth]{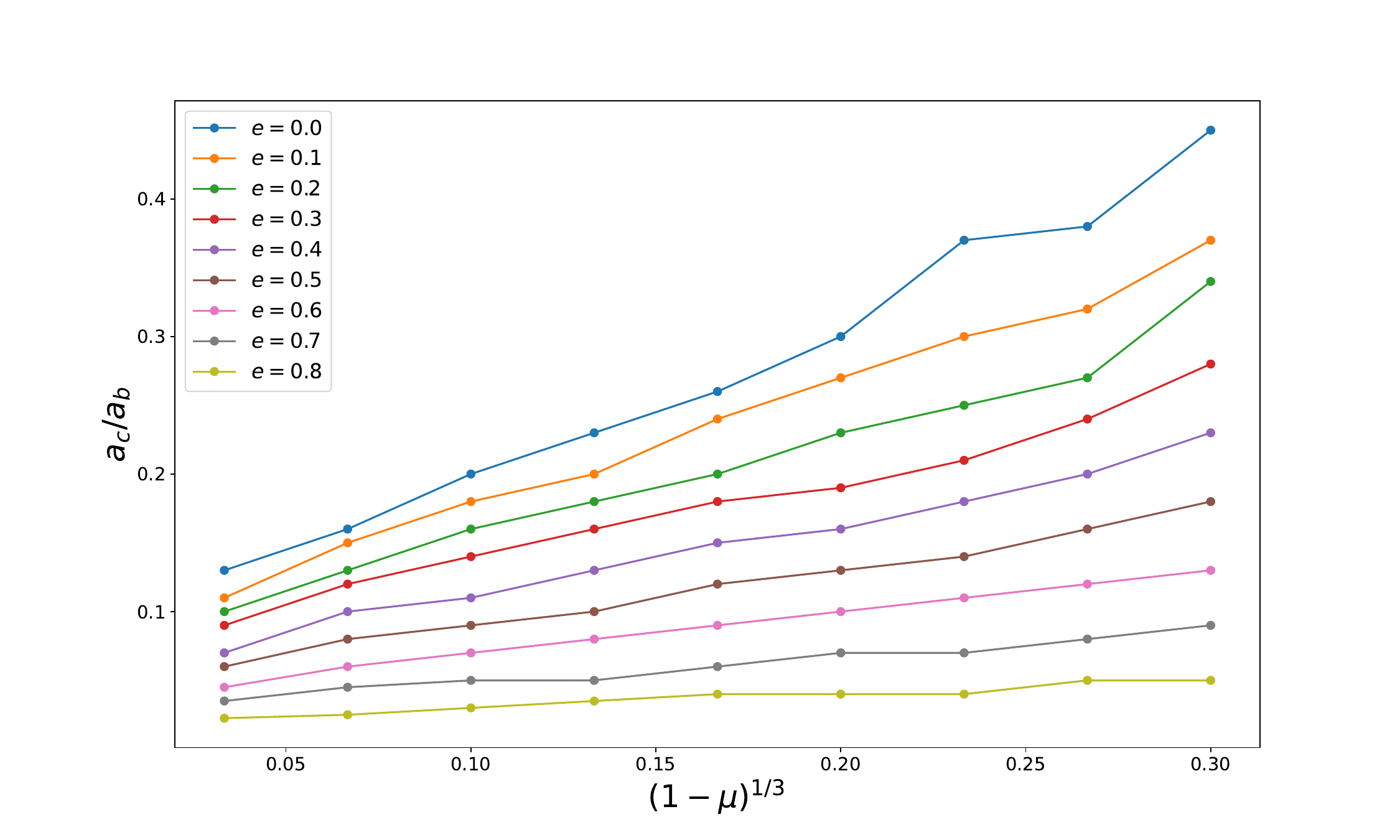}}
\caption{The relation between the ratio of the critical semi-major axis, $a_c$, for stable S-orbits in a binary system to the binary separation, $a_b$, as a function of $(1-\mu)^{1/3}$, where $\mu$ is the mass fraction of the secondary component in the binary system, $\mu = m_2 / (m_1+m_2)$. Figure based on the results from \cite{1999AJ....117..621H}.}
\label{fig:acab}
\end{figure}

\section{Binary stars affect planets}\label{Sec:Pla}
%Impact of binary stars on planet statistics
The first effect of binaries on exoplanet discoveries is the need to be cautious when we seem to detect a planetary transit. Binary stars are one of the possible contaminants, as a background eclipsing system could mimic a planetary transit. 

More importantly, binary stars may affect the existence of exoplanets. 
Planets in binary systems can either follow a S-orbits, that is, orbit one of the components (``circumprimary or circumsecondary orbits''), or P-orbits, in which case, they orbit both stars (``circumbinary planets''). 
P-orbits are generally rather far from the stars and thus harder to detect and perhaps less interesting as they are likely not in the habitable zone \citep{2020ApJ...903..141S}. Relatively few circumbinary planets are known\footnote{\url{https://exoplanet.eu/planets_binary_circum/} lists 27 such planets only, compared to 230 S-type planets.}, especially as some mechanisms during the pre-main sequence may remove such planets \citep{2018ApJ...858...86F}. An interesting system is NN Ser. Composed of a WD and a M-type companion in a 0.13-day orbit, it has been followed extensively and shows clear deviations of its eclipse timing. This led \cite{2010A&A...521L..60B} and then \cite{2014MNRAS.437..475M} to infer the presence of possibly two planets around the close binary. More recently, \cite{2023MNRAS.526.4725O} could only confirm the signal of one planet, with an orbital period of 19.5 years. Whether such planet exists and how it formed -- is it primordial or was it formed after the common envelope evolution from the ejected material? -- are still hotly debated questions.

In the case of S-orbits, they would only be stable if they are within 0.13 to 0.45 the binary orbital separation, depending on the mass ratio, and for a circular orbit. For a very eccentric orbit, these values are 0.022 and 0.05 only \citep[Fig.~\ref{fig:acab},]
[]{1999AJ....117..621H} --  see also \citet{2015ApJ...799..147J} and \citet{2020AJ....159...80Q}. Thus, it is no surprise that \cite{2014ApJ...791..111W}
``find that compared to single star systems, planets in multiple-star systems occur $4.5 \pm 3.2$, $2.6 \pm 1.0$, and $1.7 \pm 0.5$ times less frequently when a stellar companion is present at a distance of 10, 100, and 1000 au, respectively.'' More recently, 
\cite{2021MNRAS.507.3593M} find that ``binaries with a $< 1$ au fully suppress S-type planets, binaries with a = 10 au host close planets at $15^{+17}_{-12} \%$ the occurrence rate of single stars, and wide binaries with a $> 200$ au have a negligible effect on close planet formation''. They also found that about 40\% of solar-type primaries in magnitude-limited samples do not host close planets due to suppression by close stellar companions, confirming a previous result by \cite{2016AJ....152....8K}. This is likely due to the  role of stellar companions in disrupting the protoplanetary disk \citep{2010ApJ...709L.114D} within the first 1-2 Myr of disc evolution  \citep{2012ApJ...745...19K,2019ApJ...878...45B}, providing support to the disc fragmentation scenario \citep[see also][]{2020MNRAS.491.5158T}.
Binarity seems, however, to play a positive role in the existence of very massive short-period giant planets and brown dwarfs \citep{2004A&A...417..353E,2019MNRAS.485.4967F,2021MNRAS.507.3461C}. It also drives larger eccentricities of the planets  \citep{2004A&A...417..353E,2024A&A...689A.302G}.
Finally, it is also noteworthy that the orbits of transiting planets often align with the orbital planes of their stellar companions \citep{2022MNRAS.512..648D,2022AJ....163..207C,2024arXiv240510379C,2022AJ....163..160B,2023AJ....166..166L,2023AJ....165...73Z}.

\section{Binary stars and the cosmic distance scale}
Another extremely debated question is the so-called $H_0$ tension, which relies on accurately measuring the distance to the most distant objects. Such measurements are based on the cosmic distance ladder, in which different kinds of astrophysical objects probe different scales. Eclipsing binary systems offer the most  accurate way to measure the distance to the Large Magellanic Cloud, the first step on the ladder. An accuracy of 5\% is now feasible to a distance of around 3 Mpc, and so such measurements are also done in the Small Magellanic Cloud, and in the Andromeda and Triangulum galaxies \citep{2019Natur.567..200P}. 

Apart from eclipsing binaries, the first rungs on the ladder are based on cepheids. However, as these variable stars originated from stars with intermediate masses, the majority are likely to be members of binary systems \citep{2024A&A...686A.263P}. If not taken into account, this can have a drastic effect on the calibration and thus on the derived distances. Indeed, cepheids in binary systems seem to be overluminous compared to models \citep{2024ApJ...971..190E} and such binaries are now thought to be cause of the scatter in the period-luminosity relation \citep{2018ApJ...867..121G}. 

Finally, on the other end of the cosmic distance ladder are Type Ia supernovae, which are generally assumed to be syandard candles. Given that it is still unclear what is the origin of such events -- although all scenarios assume a binary model -- it is a rather amazing act of faith to rely on them for such an important endeavour; an act that clearly illustrates the importance of binary stars! 

\acknowledgements
DJ acknowledges support from the Agencia Estatal de Investigaci\'on del Ministerio de Ciencia, Innovaci\'on y Universidades (MCIU/AEI) and the European Regional Development Fund (ERDF) with reference PID-2022-136653NA-I00 (DOI:10.13039/501100011033). DJ also acknowledges support from the Agencia Estatal de Investigaci\'on del Ministerio de Ciencia, Innovaci\'on y Universidades (MCIU/AEI) and the the European Union NextGenerationEU/PRTR with reference CNS2023-143910 (DOI:10.13039/501100011033).
\bibliography{Boffin}

\end{document}